# ARTICLE

# `hklhop`: a Selection Tool for Asymmetric Reflections of Spherically Bent Crystal Analysers for High Resolution X-ray Spectroscopy

Jared E. Abramson[a], Yeu Chen[a] and Gerald T. Seidler[a]*

[a]Physics Department, University of Washington, Seattle, Washington, United States

Correspondence email: seidler@uw.edu

High resolution, hard x-ray spectroscopy at synchrotron x-ray light sources commonly uses spherically bent crystal analyzers (SBCAs) formed by shaping a single crystal wafer to a spherical backing. These Rowland circle optics are almost always used in 'symmetric' (or nearly symmetric) configurations wherein the reciprocal lattice vector used for energy selectivity via diffraction is coincident with the normal vector to the curved wafer surface. However, Gironda, et al., recently proposed that asymmetric operation of SBCA, wherein the reciprocal lattice vector is no longer normal to the wafer surface, has significant operational benefits and has been an underutilized opportunity. First, those authors find improved energy resolution through decreased Johann error, or equivalently find increased solid angle at a chosen experimental tolerance for energy broadening. Second, they find productive, high-resolution use of a large number of reciprocal lattice vectors from a single SBCA, thus enabling operation over a wide energy range without need to exchange SBCA upon making large changes in desired photon energy. These observations hold the potential to improve performance, increase flexibility and decrease cost for both laboratory and synchrotron applications. Given these motivations, we report an open-source software package, hklhop, that enables exploration of the complex space of analyzer wafer choice, experimental energy range or ranges, and desired suppression of Johann error. This package can guide both the design and the day-to-day operations of Rowland spectrometers enabled for asymmetric use.

## A Introduction

Johann-style spherically bent crystal analyzers (SBCAs) have seen extensive use at synchrotron facilities for photon-in photon-out spectroscopies[1-8] and in laboratory spectrometers for x-ray absorption fine structure and x-ray emission spectroscopy[2, 9-18]. With rare exceptions at synchrotron facilities[19], and excluding the small adjustments made for wafer miscut in laboratory instruments[20], the SBCAs are operated with the diffracting plane that provides energy resolution nominally coincident with the curved crystal surface. This 'symmetric' configuration has the source and detector at equal angles with respect to the cylindrical axis of the SBCA. However, the Rowland circle equally supports asymmetric operation, a fact previously commented on by several groups, albeit infrequently used[19, 21-23]. In asymmetric operation the diffracting crystal planes are offset by an angle $\alpha$ in the Rowland circle plane (henceforth 'Rowland plane') from the wafer surface, see Fig. 1. Note the need to introduce a distinction between the Bragg angle $\theta_B$, angle between the diffracting crystal plane and incoming x-rays, and the spectrometer or 'mechanical' angle $\theta_M$, angle between the normal crystal plane and incoming x-rays.

Recently, Gironda et. al.[24], reported the development of an SBCA-based Rowland spectrometer that is optimized for investigation of asymmetric operation, having both the necessary mechanical freedom for asymmetric source and detector positions and also a new 'clock angle', or $\varphi$, degree of freedom that rotates the SBCA about its cylindrical axis. The importance of such a rotation for bringing a desired reciprocal lattice vector into the Rowland plane was previously noted by Mortensen, et al.[20], as a one-time manual adjustment to adapt to modest wafer miscuts and thus remove the need for two-axis tilt stages for orientation of SBCAs. Gironda, et al.[24] motorize this degree of freedom enabling precise and repeatable rotations to bring any reciprocal lattice vector into the Rowland plane. For means of illustration, consider the pole plot for Si (211), Fig. 2, with $\alpha$ and $\varphi$ being the radial and azimuthal directions, respectively.

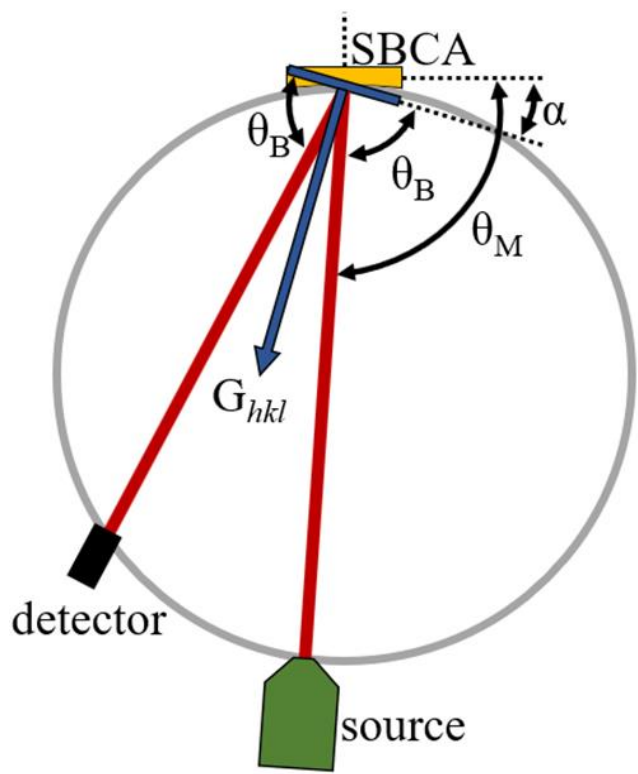

Figure 1 Diagram of asymmetric operation from the reference frame of the SBCA. $\theta_B$ is the Bragg angle, $\alpha$ is the angle between the crystal plane normal to the SBCA surface (which reflection the crystal was cut for, $G_0$) and the reflection plane being used for asymmetric operation, $G_{hkl}$, and $\theta_M$ is the angle between the SBCA surface and the incident X-rays.

While symmetric operation of such an analyzer would, allowing for harmonics, enable high resolution study of a few ~1-keV wide energy regions, there would be large gaps that in present symmetric practice would require physical exchange of SBCAs to obtain different $d$-spacings. However, as shown in Fig. 2, there are numerous reciprocal lattice vectors corresponding to different $d$-spacings within a reasonable asymmetric tilt ($\alpha$) from, for example, the Si (211) diffraction plane. Using the $\varphi$ degree of freedom the accessible energy range of the Si (211) SBCA is greatly increased.

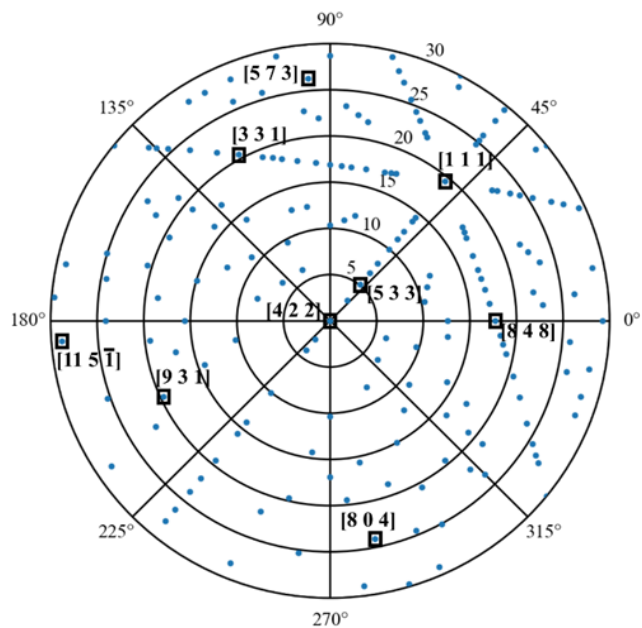

Figure 2 Pole plot demonstrating other reciprocal lattice vectors, $\vec{G}_{hkl}$, available from the Si (211) SBCA. The radial direction is $\alpha$, the angle of the reciprocal lattice vector with respect to [211], and the azimuthal direction is $\varphi$, the rotation angle of the system about the [211] direction. We show $\vec{G}_{hkl}$ with Miller indices in the range $\pm 12$, with only selected points labeled for clarity of presentation.

Gironda, et al., find two general advantages in this extended perspective on the Rowland spectrometer. First, in accordance with prior observations[25-29], but few prior implementations[30-35], they observe suppression of Johann broadening when $\theta_M$ is kept relatively near to 90 deg, even at Bragg angles where symmetric operation would require analyzer masking to prevent significantly degraded energy resolution. For example, in Fig. 3 we show both a symmetric and an asymmetric spectrometer configuration with the same SBCA that are tuned to the same energy, 8905 eV for the Cu $Kb_{1,3}$ emission line. However, the anticipated energy resolutions are not the same: symmetric operation will have large broadening from Johann error while asymmetric operation, which is close to the 'Johann normal alignment' (JNA)[36] where $\theta_M$=90 deg, will be largely immune from Johann broadening. Second, Gironda, et al.[24], in fact do find that the many reciprocal lattice vectors available from a single SBCA can often span all, or very nearly all, of the hard x-ray range where SBCAs are used. Hence, the common use of suites of SBCA to cover many different emission lines or absorption edges might be replaced with a much smaller number of SBCA, perhaps even just one, combined with the new $\varphi$ degree of freedom resulting in cost savings and in simplified operations.

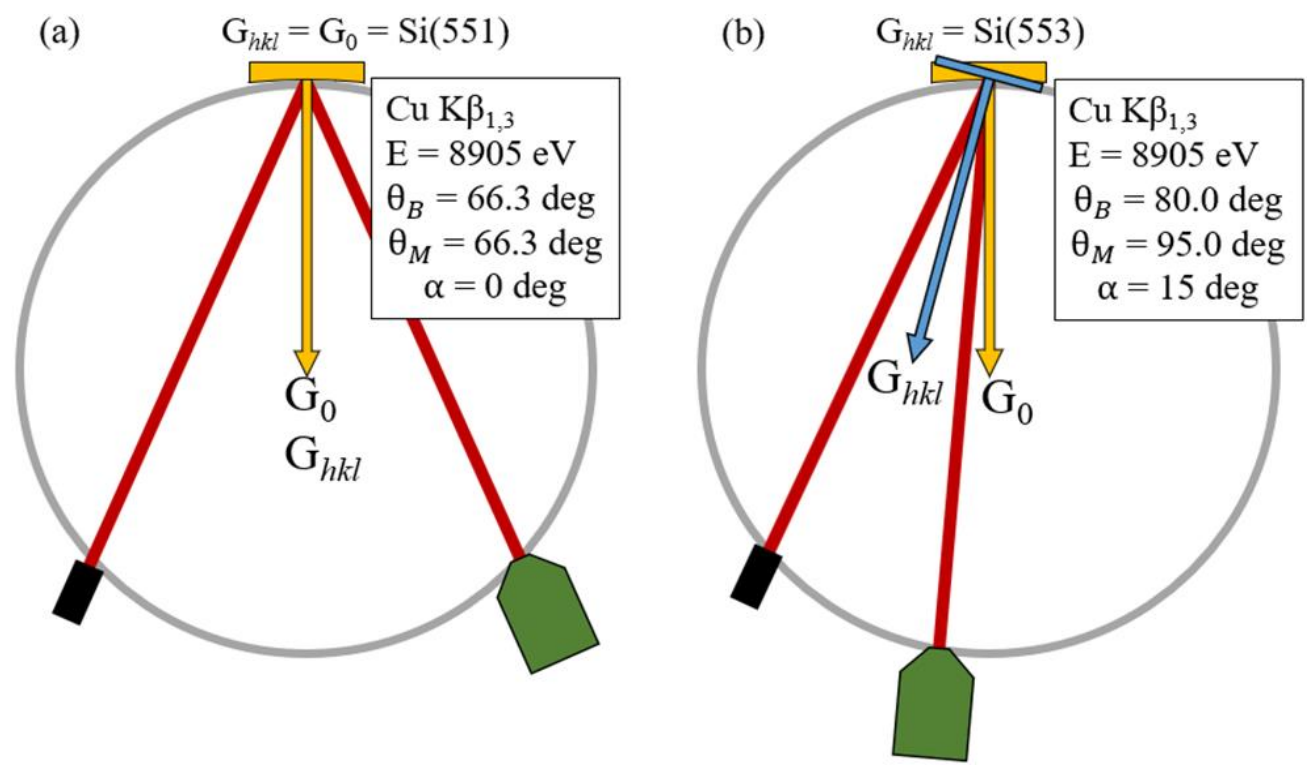

Figure 3 Two options for studying the Cu K$\beta_{1,3}$ fluorescence. (a) Symmetric operation of a Si (551) SBCA; (b) Asymmetric operation of a Si (551) SBCA using the Si (553) diffraction plane.

The above proposal to centralize asymmetric configurations in future spectrometer design and operations, however, comes with at least two limitations.  First, asymmetry degrades the in-plane focus of the analyzed radiation on the detector plane and also disadvantageously increases the sagittal defocusing due to the shorter analyzer-to-detector chord[36], possibly resulting in lost detection efficiency due to finite detector size.  This issue and the quantitative benefits in suppressing Johann error are addressed in the recent ray-tracing study of Chen, et al.[36] Second, the considerable freedom of asymmetric operation comes with a need to fully explore the combined space of analyzer wafer selection, spectroscopic reflection selection, energy ranges dictated by experiment design, and energy resolution (meaning suppression of Johann error).  We report here an open-source software package, hklhop, for this purpose.

## B Methods

The package hklhop is implemented in a Jupyter Notebook. It is available as an open source code on Github[37]. Ray tracing calculations are geometrical, with no allowance for strain effects[2], using the xrt package [38] and following the methods of Chen et. al.[36]

## C Results and Discussion

The package, schematically shown in Fig. 4, was developed with one main function, `hkl_selection`, for investigating the possible *hkl* space of specified SBCAs and energies and another function, `sbca_selection`, for investigating the possible *hkl* space when only an energy is specified (no limit to SBCA wafer orientation). Here we consider five workflows that follow from realistic experimental needs: the user has a single SBCA and target energy to

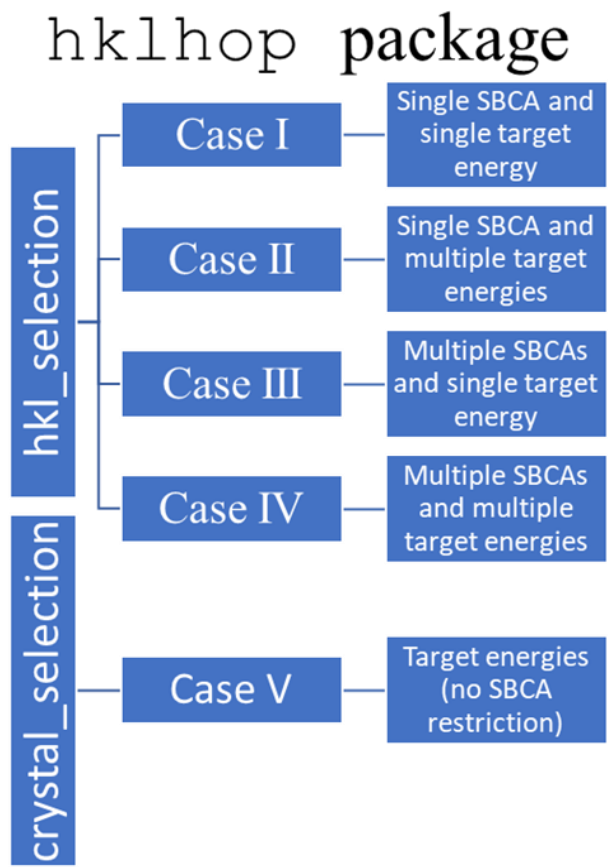

Figure 5 Descriptions of the five main use cases for the `hklhop` package. The two main functions in the package are `hkl_selection` and `crystal_selection`.

Table for target energy of 8639 eV

| | $G_0$ | $G_{hkl}$ | $\theta_B$ (deg) | $\alpha$ (deg) | $\theta_M$ (deg) | $G_{\phi=0}$ | $\phi$ (deg) |
|---|---|---|---|---|---|---|---|
| 0 | Si [5 5 1] | [6 4 2] | 81.48 | 13.34 | 94.82 | [1 0 0] | 42.97 |
| 1 | Si [5 5 1] | [6 4 -2] | 81.48 | 26.08 | 107.56 | [1 0 0] | -56.57 |
| 2 | Si [5 5 1] | [5 5 1] | 70.70 | 0.00 | 70.70 | [1 0 0] | 22.01 |
| 3 | Si [5 5 1] | [5 5 -1] | 70.70 | 16.10 | 86.80 | [1 0 0] | -82.03 |
| 4 | Si [5 5 1] | [7 1 1] | 70.70 | 36.49 | 107.19 | [1 0 0] | 10.64 |
| 5 | Si [5 5 1] | [4 4 4] | 66.29 | 27.21 | 93.51 | [1 0 0] | 97.97 |
| 6 | Si [5 5 1] | [4 4 -4] | 66.29 | 43.31 | 109.61 | [1 0 0] | -82.03 |
| 7 | Si [5 5 1] | [5 3 3] | 60.07 | 23.33 | 83.40 | [1 0 0] | 64.98 |
| 8 | Si [5 5 1] | [5 3 -3] | 60.07 | 37.81 | 97.87 | [1 0 0] | -61.43 |

Figure 4 Output from the `hkl_selection` function for Case I, having a single SBCA and single target energy, studying Zn K$\alpha$ with a Si (551) SBCA. The green box has been added for indication of the optimal reciprocal lattice vector option.

study (Case I); the user has a single SBCA and multiple energies to study (Case II); the user has multiple SBCAs and a single energy to study (Case III); the user has multiple SBCAs and multiple energies to study (Case IV); and the user has just an energy to study with no restriction on SBCA (Case V), see Figure 4. For Case I-IV the user must input a list of SBCAs (presently restricted to Si or Ge), a list of target energies, and constraints on possible asymmetric orientations. These restrictions include the maximum index of the diffracting plane, the $\theta_B$ range, the $\theta_M$ range, and the necessary energy range above and below the target energy. For Case V the user instead inputs the maximum index of possible SBCA options, a target energy and the same restrictions

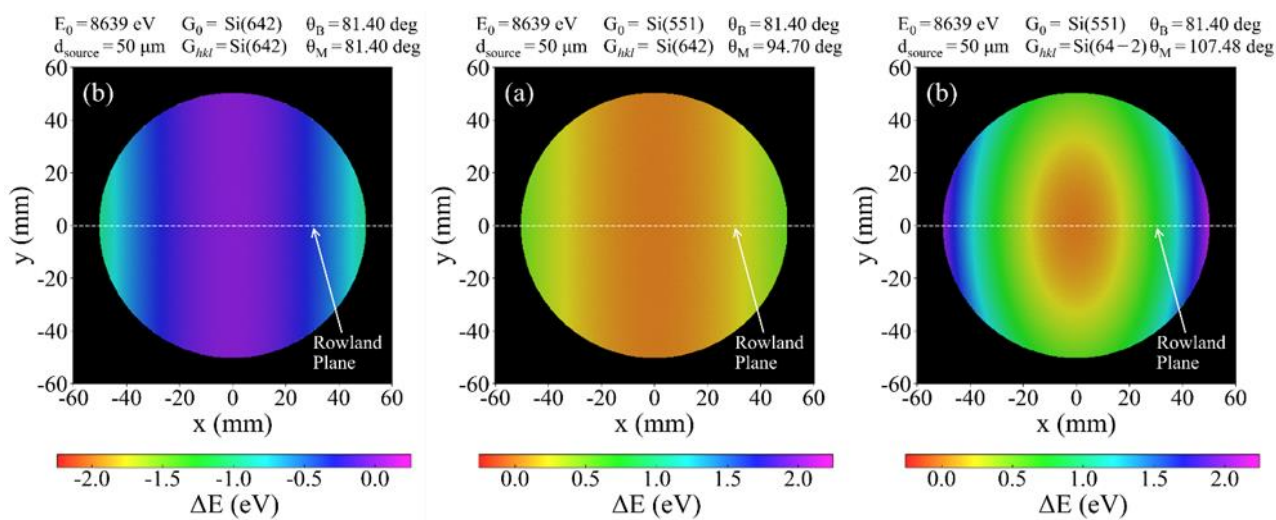

Figure 6 Predicted energy dispersion across the SBCA face for Si (642) reflections from symmetric and two asymmetric configurations: (a) using the (642) $G_{hkl}$ symmetrically having $\theta_B$ and $\theta_M$ of 81.40 deg (b) using the (642) $G_{hkl}$ from a Si (551) SBCA having $\theta_B$ of 81.40 deg and $\theta_M$ of 94.70 deg (c) using the (64-2) $G_{hkl}$ from a Si (551) SBCA having $\theta_B$ of 81.40 deg and $\theta_M$ of 107.48 deg. The larger broadening seen for the (642) symmetric and (64-2) asymmetric configuration is due to the inferior $\theta_M$.

as for Case I-IV. Below, we provide representative studies for each of the five workflows.

For Case I, where the user has a single SBCA and single energy to study, Fig. 5 shows the output with a Si (551) SBCA to isolate Zn Kα at 8639 eV, such as is needed for a high-energy resolution fluorescence detection (HERFD) study[39, 40]. Generally, Si (642) G$_{hkl}$ is ideal for symmetric operation for Zn Kα as it has the best $\theta_B$ = 81.48 deg; given symmetric operation, $\theta_M = \theta_M$. While 81 deg is generally favorable, accessing the (642) crystal plane asymmetrically from the Si (551) SBCA results in the same Bragg angle but $\theta_M$ is now 94.82 deg, closer to the JNA and consequently leading to better energy resolution via decreased Johann error. This is further demonstrated by comparing the (642) and (64-2) G$_{hkl}$ from Fig. 5. In Fig. 6 we show ray tracing results for the energy response function mapped across the face of the SBCA for these reflections. Note that the symmetric Si (642) shows 1-eV error at the SBCA edge while the asymmetric Si (642) shows 0.5-eV error and the asymmetric Si (64-2) shows 2-eV error. This difference between the three diffraction planes leads to a choice of the asymmetric (642) G$_{hkl}$ over the symmetric (642) and asymmetric (64-2) G$_{hkl}$ reflections.

The choice of which G$_{hkl}$ to use depends largely on experimental constraints but typically the most important criteria is $\theta_B$, with larger angles being favorable[25-29]. This is followed in importance by $\theta_M$, picking the reflection closest to JNA, and with no other considerations besides energy resolution the (642) G$_{hkl}$ with the Si (551) SBCA would be the best choice for studying Zn Kα in a laboratory setting, see the green box in Fig. 5.

Next, in Case II the user has a single SBCA and multiple energies to study. An example output is displayed in Fig. 7 for a Ge (620) SBCA being used to study the Mn Kα (5899 eV), Co Kα (6930 eV), and Ni Kα (7478 eV) fluorescence. The tables, generated separately for each energy, demonstrate the flexibility of working asymmetrically in allowing one SBCA to access multiple energies while retaining good resolution.

Table for target energy of 5899 eV

| | $G_0$ | $G_{hkl}$ | $\theta_B$ (deg) | $\alpha$ (deg) | $\theta_M$ (deg) | $G_{\phi=0}$ | $\phi$ (deg) |
|---|---|---|---|---|---|---|---|
| 0 | Ge [6 2 0] | [5 1 1] | 74.82 | 13.16 | 87.98 | [1 0 0] | 57.69 |
| 1 | Ge [6 2 0] | [5 1 -1] | 74.82 | 13.16 | 87.98 | [1 0 0] | -57.69 |
| 2 | Ge [6 2 0] | [5 -1 -1] | 74.82 | 31.57 | 106.39 | [1 0 0] | -21.57 |
| 3 | Ge [6 2 0] | [4 2 2] | 65.49 | 25.35 | 90.85 | [1 0 0] | 107.55 |
| 4 | Ge [6 2 0] | [4 2 -2] | 65.49 | 25.35 | 90.85 | [1 0 0] | -107.55 |

Table for target energy of 6930 eV

| | $G_0$ | $G_{hkl}$ | $\theta_B$ (deg) | $\alpha$ (deg) | $\theta_M$ (deg) | $G_{\phi=0}$ | $\phi$ (deg) |
|---|---|---|---|---|---|---|---|
| 0 | Ge [6 2 0] | [5 3 1] | 69.28 | 15.82 | 85.1 | [1 0 0] | 141.67 |
| 1 | Ge [6 2 0] | [5 3 -1] | 69.28 | 15.82 | 85.1 | [1 0 0] | -141.67 |
| 2 | Ge [6 2 0] | [5 1 -3] | 69.28 | 31.21 | 100.5 | [1 0 0] | -78.10 |

Table for target energy of 7478 eV

| | $G_0$ | $G_{hkl}$ | $\theta_B$ (deg) | $\alpha$ (deg) | $\theta_M$ (deg) | $G_{\phi=0}$ | $\phi$ (deg) |
|---|---|---|---|---|---|---|---|
| 0 | Ge [6 2 0] | [5 3 3] | 73.90 | 29.77 | 103.67 | [1 0 0] | 112.86 |
| 1 | Ge [6 2 0] | [5 3 -3] | 73.90 | 29.77 | 103.67 | [1 0 0] | -112.86 |
| 2 | Ge [6 2 0] | [6 0 -2] | 67.92 | 25.84 | 93.76 | [1 0 0] | -46.51 |

Figure 7 Output from `hkl_selection` function for Case II, having a single SBCA and multiple target energies, studying Ni Kα, Mn Kα, and Co Kα all with a Ge (620) SBCA. The green boxes have been added to indicate the optimal choice of reciprocal lattice vector. The case shows the ability of a single SBCA to cover a range of energies.

Using the same diffraction plane selection method as for Case I we see that Ge (511) or (55-1), Ge (531) or (53-1), and Ge (533) or (53-3) reflection planes will be used, respectively, for Mn Kα, Co Kα, and Ni Kα fluorescence, see the green boxes in Fig. 7. This selection method optimizes energy resolution but neglects the effects of asymmetric operation on beam focusing. As operation gets further from a symmetric case, i.e. $\alpha$ increases, the sagittal focusing gets worse for constant $\theta_B$ decreasing the detected signal. Therefore, if signal is very important $\alpha$ should be prioritized above $\theta_M$ as long as $\theta_M$ is still a reasonable value. Further discussion can be found in Chen et. al.[36]

To further examine the flexibility of working asymmetrically, Fig. 8 shows the full coverage of unique asymmetric reflections (those that do not access the exact same energy range) from Si (100), Si (211), Si (551), and Si (642) SBCAs within an energy range of 3750 to 10000 eV. These plots are generated by an auxiliary function of the package, bar_chart. Due to the higher symmetry for the Si (100) SBCA there are fewer crystal planes with unique *d*-spacing within reasonable α values, leading to poorer energy range coverage and therefore a preference for low symmetry SBCAs for asymmetric operation. At lower energies there are fewer crystal planes with correct *d*-spacings, leading to sparser energy range coverage for all SBCAs, high or low symmetry.

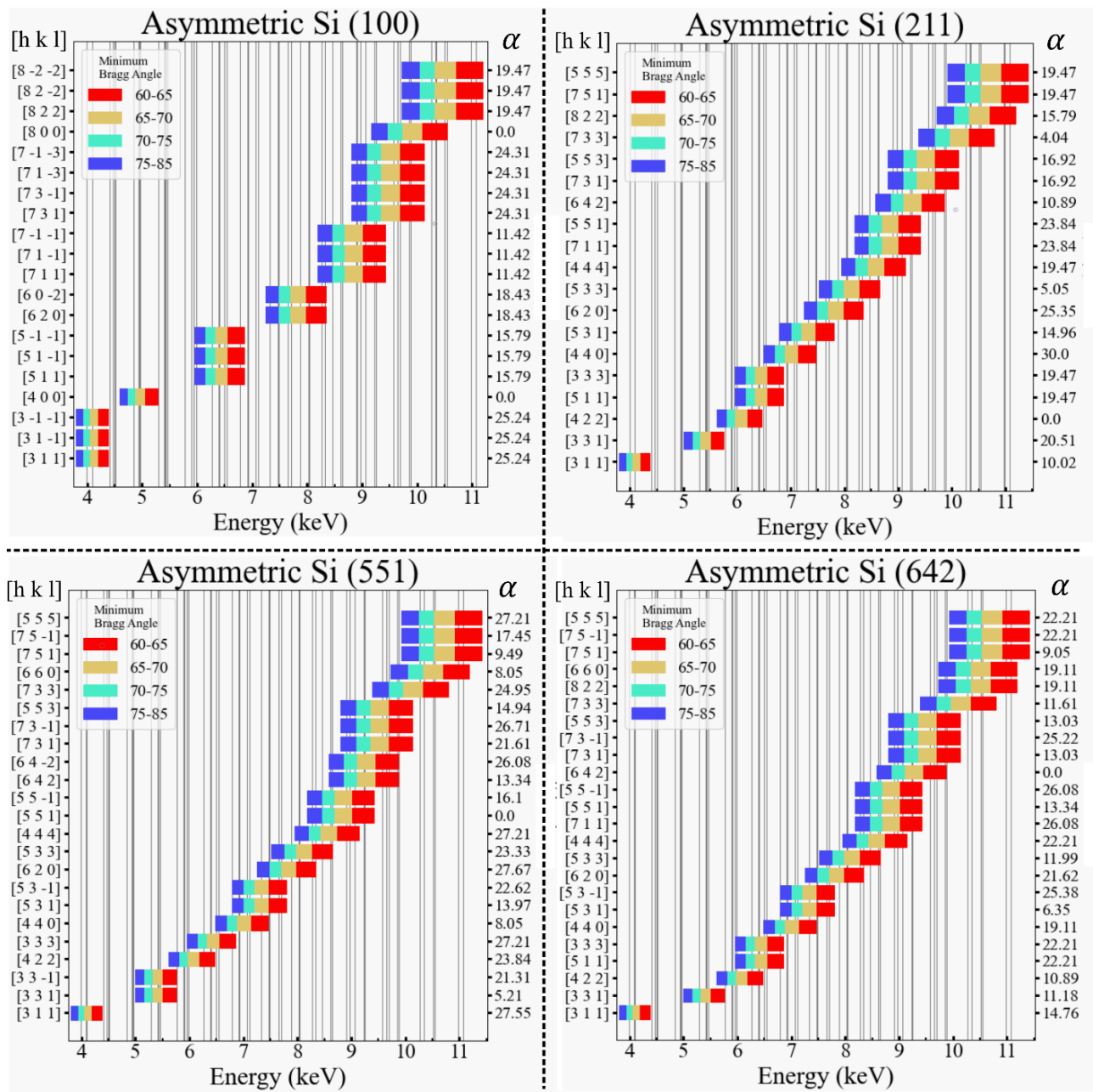

Figure 8 Bar charts generated by an auxiliary function of the package, bar_chart, displaying the energy ranges of the Si (100), Si (211), Si (551), and Si (642) SBCAs when used asymmetrically. The y-axis of each graph shows accessible crystal planes for ~4-10 keV.

For Case III with a single energy and multiple SBCAs, Fig. 9 gives output for the energy of the Pt VtC fluorescence with either a Si (111), Si (211), or Si (642) SBCA. Comparing $G_{hkl}$ across all three SCBAs the optimal configuration to study Pt VtC can be chosen as before: the [931] reciprocal lattice vector accessed with the Si (642) SBCA has $\theta_M$ closest to JNA for the options with the best $\theta_B$, see the green box in Fig. 9. One

---

Table for target energy of 11070 eV

|    | $G_0$    | $G_{hkl}$ | $\theta_B$ (deg) | $\alpha$ (deg) | $\theta_M$ (deg) | $G_{\phi=0}$ | $\phi$ (deg) |
|----|----------|-----------|------------------|----------------|------------------|--------------|--------------|
| 0  | Si [6 4 2] | [9 3 1]  | 79.69            | 17.72          | 97.40            | [1 0 0]      | -8.86        |
| 1  | Si [2 1 1] | [9 3 1]  | 79.69            | 19.69          | 99.38            | [1 0 0]      | -26.10       |
| 2  | Si [6 4 2] | [9 3 -1] | 79.69            | 26.29          | 105.98           | [1 0 0]      | -31.95       |
| 3  | Si [1 1 1] | [6 6 4]  | 75.35            | 10.02          | 85.38            | [1 0 0]      | -60.00       |
| 4  | Si [6 4 2] | [6 6 4]  | 75.35            | 14.38          | 89.73            | [1 0 0]      | 157.43       |
| 5  | Si [2 1 1] | [6 6 4]  | 75.35            | 16.78          | 92.13            | [1 0 0]      | -148.52      |
| 6  | Si [6 4 2] | [7 5 3]  | 69.98            | 4.12           | 74.11            | [1 0 0]      | 136.91       |
| 7  | Si [2 1 1] | [7 5 3]  | 69.98            | 9.65           | 79.64            | [1 0 0]      | -112.21      |
| 8  | Si [1 1 1] | [7 5 3]  | 69.98            | 18.09          | 88.07            | [1 0 0]      | -30.00       |
| 9  | Si [2 1 1] | [9 1 1]  | 69.98            | 26.33          | 96.32            | [1 0 0]      | 0.00         |
| 10 | Si [6 4 2] | [9 1 1]  | 69.98            | 28.35          | 98.33            | [1 0 0]      | 5.93         |
| 11 | Si [6 4 2] | [9 1 -1] | 69.98            | 34.77          | 104.76           | [1 0 0]      | -14.96       |
| 12 | Si [6 4 2] | [7 5 -3] | 69.98            | 34.77          | 104.76           | [1 0 0]      | -71.21       |
| 13 | Si [2 1 1] | [9 1 -1] | 69.98            | 36.24          | 106.22           | [1 0 0]      | -15.23       |
| 14 | Si [6 4 2] | [8 4 0]  | 67.29            | 17.02          | 84.31            | [1 0 0]      | -43.09       |
| 15 | Si [2 1 1] | [8 4 0]  | 67.29            | 24.09          | 91.38            | [1 0 0]      | -50.77       |
| 16 | Si [1 1 1] | [8 4 0]  | 67.29            | 39.23          | 106.52           | [1 0 0]      | -30.00       |

Figure 9 Output from `hkl_selection` function for Case III, having multiple SBCAs and a single target energy, studying Pt VtC fluorescence with a Si (111), Si (211), or Si (642) SBCA. The blue and green boxes have been added to highlight the optimal choices of reciprocal lattice vector. The green box is for a large source size spectrometer, e.g. laboratory, where $\theta_B$ is the most important parameter for resolution, whereas the blue box is for a small source size spectrometer, e.g. micro-focused beamline, where $\theta_M$ is the most important parameter for resolution.

common experimental constraint that will change this choice is having a small source size, such as from a micro-focused synchrotron beam. For small source sizes, the value of $\theta_M$ controls analyzer selection over $\theta_B$ because energy broadening becomes dominated by Johann error rather than source size, with the caveat that a good $\theta_B$ is still beneficial to minimize the consequences of off-circle alignment. Hence, for a small source size the [664] reciprocal lattice vector accessed with the Si (642) SBCA is preferred, see the blue box in Fig. 9.

For Case IV, investigating the use of multiple SBCAs to study multiple energies, Fig. 10 considers measurement of V valence-to-core (VtC) and Fe Kβ fluorescence using either a Si (100), Si (311), or Si (533) SBCA. By combining the multiple target energies and multiple SBCA options from Case II and Case III the user

---

Table for target energy of 5463 eV

|   | $G_0$    | $G_{hkl}$ | $\theta_B$ (deg) | $\alpha$ (deg) | $\theta_M$ (deg) | $G_{\phi=0}$ | $\phi$ (deg) |
|---|----------|-----------|------------------|----------------|------------------|--------------|--------------|
| 0 | Si [5 3 3] | [3 3 1]  | 65.64            | 19.16          | 84.80            | [1 0 0]      | -98.67       |
| 1 | Si [3 1 1] | [3 3 1]  | 65.64            | 25.94          | 91.58            | [1 0 0]      | -132.13      |
| 2 | Si [3 1 1] | [3 3 -1] | 65.64            | 40.46          | 106.10           | [1 0 0]      | -90.00       |
| 3 | Si [5 3 3] | [3 3 -1] | 65.64            | 42.72          | 108.36           | [1 0 0]      | -73.04       |

---

Table for target energy of 7058 eV

|   | $G_0$    | $G_{hkl}$ | $\theta_B$ (deg) | $\alpha$ (deg) | $\theta_M$ (deg) | $G_{\phi=0}$ | $\phi$ (deg) |
|---|----------|-----------|------------------|----------------|------------------|--------------|--------------|
| 0 | Si [3 1 1] | [5 3 1]  | 73.13            | 14.46          | 87.59            | [1 0 0]      | -106.78      |
| 1 | Si [5 3 3] | [5 3 1]  | 73.13            | 17.49          | 90.63            | [1 0 0]      | -52.67       |
| 2 | Si [3 1 1] | [5 3 -1] | 73.13            | 29.96          | 103.09           | [1 0 0]      | -73.22       |
| 3 | Si [1 0 0] | [5 3 1]  | 73.13            | 32.31          | 105.45           | [0 0 1]      | 71.57        |
| 4 | Si [1 0 0] | [5 3 -1] | 73.13            | 32.31          | 105.45           | [0 0 1]      | 108.43       |
| 5 | Si [1 0 0] | [5 1 -3] | 73.13            | 32.31          | 105.45           | [0 0 1]      | 161.57       |
| 6 | Si [1 0 0] | [5 -1 -3]| 73.13            | 32.31          | 105.45           | [0 0 1]      | -161.57      |
| 7 | Si [5 3 3] | [4 4 0]  | 66.21            | 30.38          | 96.60            | [1 0 0]      | -81.33       |
| 8 | Si [3 1 1] | [4 4 0]  | 66.21            | 31.48          | 97.70            | [1 0 0]      | -106.78      |

Figure 10 Output from `hkl_selection` function for Case IV, having multiple SBCAs and multiple target energies. Here V VtC and Fe Kβ fluorescence is being studied by a Si (100), Si (311), or Si (533) SBCA. The green boxes have been added to indicate the optimal choice of reciprocal lattice vector for each energy.

can investigate the full range of options available to them and choose the best SCBA and $G_{hkl}$ combination at each energy or choose the best single SBCA that can study all the energies having $G_{hkl}$ options for each. In this limited example of three SBCAs and two energies, all SBCAs can reach reciprocal lattice vectors with the best $\theta_B$ but having different α, and therefore differing $\theta_M$. Optimizing $\theta_M$ for the best $\theta_B$ leads to a choice of Si (311) SBCA for V VtC and Si (533) for Fe Kβ, boxed in green in Fig. 10. If a single SBCA was desired to measure both V VtC and Fe Kβ the Si (311) SBCA has the best $\theta_M$ across the two energies and would be chosen.

Finally, for Case V, where the user has selected an energy but is open to all SBCA choices, we consider U Lα$_1$ in Fig. 11 which demonstrates typical output of the sbca_selection function. From this table an SBCA can be chosen to study a fluorescence line at an optimal configuration, see the green box in Fig. 11 for

```
--------------------------------------------------
Table for target energy of 13615 eV

       G₀         G_hkl    θ_B (deg)  α (deg)   θ_M (deg)  G_φ=0    φ (deg)
 0   Si [3 1 1]   [11 3 3]   81.36     4.15      85.51    [1 0 0]    0.00
 1   Si [5 3 3]   [9 7 3]    81.36    14.04      95.40    [1 0 0]  -81.33
 2   Si [3 1 1]   [9 7 3]    81.36    18.87     100.23    [1 0 0] -132.13
 3   Si [5 3 3]   [11 3 3]   81.36    19.22     100.58    [1 0 0]    0.00
 4   Si [1 0 0]   [11 3 3]   81.36    21.09     102.45    [0 0 1]   45.00
 5   Si [1 0 0]   [11 3 -3]  81.36    21.09     102.45    [0 0 1]  135.00
 6   Si [1 0 0]   [11 -3 -3] 81.36    21.09     102.45    [0 0 1] -135.00
 7   Si [5 3 3]   [8 6 6]    77.94     6.37      84.31    [1 0 0]  180.00
 8   Si [3 1 1]   [10 6 0]   77.94    21.45      99.38    [1 0 0]  -84.26
 9   Si [3 1 1]   [8 6 6]    77.94    21.45      99.38    [1 0 0]  180.00
10   Si [5 3 3]   [10 6 0]   77.94    27.23     105.16    [1 0 0]  -52.67
11   Si [1 0 0]   [10 6 0]   77.94    30.96     108.90    [0 0 1]   90.00
12   Si [1 0 0]   [10 0 -6]  77.94    30.96     108.90    [0 0 1] -180.00
--------------------------------------------------
```

Figure 11 Output from `sbca_selection` function for Case V, having just a target energy, 13,615 eV for U L$\alpha_1$. The function finds all reciprocal lattice vectors, $G_{hkl}$, that access the energy above a certain $\theta_B$ and find the SBCA reciprocal lattice vectors, $G_0$, that reach it and the associated $\alpha$ and $\theta_M$.

this example. While only one energy is shown in Fig. 11 the sbca_selection function accepts input of a list of energies, making a table for each one. This can help plan experiments where the number of SBCA changes can be minimized.

## D Conclusions

Asymmetric Rowland circle geometries greatly improve the energy range of SBCAs while also decreasing Johann error but have been historically underutilized due to lack of spectrometer capability, knowledge of the benefits, and guidance on how to operate asymmetrically. The hklhop package provides users with the missing guidance with tools to choose the best combinations of analyzer and reflection for any given experiment. Taking input tailored to a specific experimental set up, users are provided with tables of analyzers, reflections, and important parameters which aid in the choice and implementation of asymmetric operation. While the package is currently limited to Si and Ge SBCA, other analyzer materials can be easily added.

## Acknowledgements

The project was supported by U.S. Department of Energy in the Nuclear Energy University Program under Contract No. DE-NE0009158